\documentclass[12pt,a4paper]{article}

\def\ben{\begin{enumerate}} \def\een{\end{enumerate}}
\def\beq{\begin{equation}} \def\eeq{\end{equation}}
\def\beqn{\begin{equation*}} \def\eeqn{\end{equation*}}
\def\bea{\begin{eqnarray}} \def\eea{\end{eqnarray}}
\def\ba{\begin{array}} \def\ea{\end{array}}
\def\beann{\begin{eqnarray*}} \def\eeann{\end{eqnarray*}}
\def\beasn{\begin{sneqnarray}} \def\eeasn{\end{sneqnarray}}
\def\bi{\begin{itemize}} \def\ei{\end{itemize}}
\def\be{\begin{enumerate}} \def\ee{\end{enumerate}}

\def\u{\underline}

\def\u{\"u}
\def\o{\"o}
\def\ea{\'e}

\def\a{\"a}
\def\mf{\mathfrak}

\usepackage{graphicx}
\usepackage{xcolor}
\usepackage[paper=a4paper,left=30mm,right=30mm,top=30mm,bottom=30mm]{geometry}
\usepackage{amsmath}
\usepackage{amsthm}
\usepackage{amssymb}
\usepackage{eufrak}

\numberwithin{equation}{section}
\title{Toward a quantum theory of gravity: Syracuse 1949-1962}

\author
{Donald Salisbury$^{1,2}$\\
\\
\normalsize{$^{1}$Austin College, 900 North Grand Ave, Sherman, Texas 75090, USA}\\
\normalsize{$^{2}$Max Planck Institute for the History of Science,}\\
\normalsize{Boltzmannstrasse 22, 14195 Berlin, Germany}}

\begin{document} 

\maketitle

\begin{abstract}
Peter Bergmann and his students embarked in 1949 on a mainly canonical quantization program whose aim was to take into account the underlying four-dimensional diffeomorphism symmetry in the transition from a Lagrangian to a Hamiltonian formulation of Einstein's theory. Along the way they developed techniques for dealing in phase space with the arbitrariness that arises in classical solutions given data at a given coordinate time. They argued that even though one seemed to destroy the full covariance through the focus on a temporal foliation of spacetime, this loss was illusory. In work undertaken with Anderson and I. Goldberg in the early 1950's they constructed explicit expressions for both the generator of transformations between solutions which could be physically distinct, i.e., did not necessarily correspond to changes in spacetime coordinates, and invariant transformations that did correspond to four-dimensional diffeomorphisms. They argued that the resulting factor algebra represented diffeomorphism invariants that were the correct candidates for promotion to quantum operators. Early on they convinced themselves that only the corresponding construction of classical invariants could adequately reflect the fully relativistic absence of physical meaning of spacetime coordinates. Efforts were made by Bergmann students Newman and Janis to construct these classical invariants. Then in the late 1950's Bergmann and Komar proposed a comprehensive program in which classical invariants could be constructed using the spacetime geometry itself to fix intrinsic spacetime landmarks. At roughly the same time Dirac formulated a new criterion for identifying initial phase space variables, one of whose consequences was that Bergmann himself abandoned the gravitational lapse and shift as canonical variables. Furthermore, Bergmann in 1962 interpreted the Dirac formalism as altering the very nature of diffeomorphism symmetry. One class of infinitesimal diffeomorphism was to be understood as depending on the perpendicular to the given temporal foliation. Thus even within the Bergmann school program the preservation of the full four-dimensional symmetry in the Hamiltonian program became problematic. Indeed, the ADM and associated Wheeler-DeWitt program that gained and has retained prominence since this time abandoned the full symmetry. There do remain dissenters - raising the question whether the field of quantum gravity has witnessed a Renaissance in the ensuing decades - or might the full four-dimensional symmetry yet be reborn?

\end{abstract}

\section{Introduction}

Peter G. Bergmann served as a research assistant to Albert Einstein at the Institute for Advanced Study in Princeton from 1936 to 1941, immediately after receiving his doctoral degree at Charles University in Prague. While still working on his dissertation in which he dealt with the quantum description of a harmonic oscillator in a curved space, he wrote to Einstein that `As you can see, my training in any direction is in no way to be understood as complete. It is clear to me that I have much to learn in the field of relativity theory as well as quantum mechanics. If it were possible I would gladly continue to work in the search for a link between these two fields."\footnote{"Wie Sie ersehen k\o nnen, ist meine Ausbildung noch keineswegs in ergendeiner Richtung als abgeschlossen zu betrachten. Ich bin mir dar\u ber klar, dass ich besonders auf dem Gebiet der Relativit\a stheorie einerseits, der Quantentheorie anderseits, noch sehr viel zu lernen habe. Wenn es mir m\o glich w\a re , w\u rde ich sehr gern in der Richtung weiterarbeiten, die Verbindung zwishen diesen beiden Gebieten zu suchen." The Albert Einstein Archives at the Hebrew University of Jerusalem (AEA), 6-222, Letter from Bergmann to Einstein dated 14 March, 1936. } His work with Einstein was mainly concerned with unified field theory, but he continued to think about this problem. Having secured a tenure-track position at Syracuse University in 1947, following war-related research from 1944-47, he was finally able to focus on the problem that would occupy him for the remainder of his career. His first groundbreaking paper dealt with a generalized approach to non-linear field theories like general relativity, with an eye toward quantization \cite{Bergmann:1949aa}. Einstein at the time did not seem to endorse this enterprise. In response to a note from Bergmann\footnote{AEA, 6-282, Letter from Bergmann to Einstein dated January 24, 1949} expressing a hope that they might get together to discuss these ideas he wrote ``You are seeking an independent new path toward a solution of the fundamental problems. No one can help in this effort, and least of all someone who to some extent has his own fixed ideas. You know, for example, that on the basis certain considerations I firmly believe that the probability concept can not play a primary role in the description of reality. You seem to believe that one should first set up a field theory and then subsequently `quantize' it \ldots Your attempt to carry out an abstract field theory without knowing in advance the formal nature of the field quantities seems to me unfortunate, being formally deficient and indeterminate.\footnote{``Sie suchen einen selbst\a ndigen und neuen Weg zur L\o sung der prizipiellen Schwierigkeiten. Bei diesem Bestreben kann einem Niemand helfen, am wenigsten Einer, der einigermassen fixierte Ideen hat. Sie wissen z. B., dass ich auf Grund gewisser Ueberlegungen fest glaube, dass der Wahrscheinlichkeitsbegriff nicht prim\a r in die Realit\a tsbeschreibung eingehen darf, w\a hrend Sie daran zu glauben scheinen, dass man zuerst eine Feldtheorie aufzustellen und diese dann nachtr\a glich zu `quantisieren' hat \ldots Ihr Versuch, eine Feldtheorie abstrakt durchzuf\u hren, ohne von vornherein \u ber die formale Natur der Feldgr\o ssen zu verf\u gen, erscheit mir nicht gl\u cklich, weil formal zu arm und unbestimmt." AEA, 6-282, Letter from Einstein to Bergmann dated January 26, 1949.} Nevertheless, Einstein did in 1954 recommend approval of Bergmann's application for National Science Foundation funding of his project entitled Quantum Theory of Gravitation with the words ``All physicists are convinced of the high truth value of the probabilistic quantum theory and of the general relativity theory \ldots There are presently only few theoretical physicists who have penetrated deeply enough into both theories to be able to undertake such an attempt at all. Dr. Bergmann is one of the few who are completely at home in both theories."\footnote{AEA, 6313, Recommendation dated April 18, 1954.} Einstein was, however, still not willing to participate in this effort according to Joshua Goldberg. Following Goldberg's brief meeting with Einstein in 1954, Bergmann contacted Einstein to inquire whether the three of them might collaborate. Einstein declined. See \cite{Salisbury:2020ab}.

 I will focus in this essay on the work of Bergmann and his collaborators from 1949 to 1962, where appropriate comparing and contrasting with contemporary work in the `opposing' camps. As far as possible I will try to structure this essay in the chronological order in which problems were addressed by the Syracuse associates. The plan was to eventually develop a Hamiltonian formalism and then to undertake a canonical quantization. The fundamental challenge was to properly take into account the symmetry of Einstein's theory under general coordinate transformations. The profound consequence from Bergmann's perspective was that spacetime coordinates were  themselves devoid of physical content. He firmly believed that the ultimate quantum theory must deal exclusively with operators that were invariant under the action of this symmetry group. This view is in sharp contrast to the geometrodynamic principles that were proposed by Wheeler at roughly the time at which my narrative terminates. Bergmann may well have contributed to the dominance of the $3+1$ formalism with his interpretation of Dirac's 1958 variables, published in his 1962 overview of general relativity.
 
  In Section 2, after addressing the first foray by Bergmann (B49)  \cite{Bergmann:1949aa} and Bergmann Brunings (BB49)  \cite{Bergmann:1949aa,Bergmann:1949ab} that constituted the theoretical basis of much that was to follow, I turn in 1950 to Bergmann and his students Penfield, Schiller, and Zatzkis (BPSZ50) \cite{Bergmann:1950ab} who constructed a Hamiltonian for Einstein-Maxwell theory in a formalism in which the spacetime coordinates evolved in an arbitrary parameter time. After the discovery by Penfield (P50) \cite{Penfield:1950aa} that the parameterized formalism could be abandoned and that four of the Einstein equations arose from the demand that primary constraints be preserved in time, Anderson and Bergmann (AB51) in 1951 \cite{Anderson:1951aa} derived secondary constraints in a generic diffeomorphism covariant theory and wrote down the phase space generator of general coordinate transformations. These studies was undertaken before the Syracuse group became aware of the earlier pioneering work of L\'eon Rosenfeld (R30) \cite{Rosenfeld:1930aa}, and they became aware of Dirac's new approach to constrained dynamics only when BPSZ50 was already a preprint. Pirani and Schild had in the meantime effectively simultaneously derived their own Hamiltonian employing Dirac's method, also in the parameterized version. I will show in Schiller's thesis (S53) \cite{Schiller:1952aa} that Rosenfeld had strongly influenced him. 
 
  In Section 3, after a brief look at Anderson's 1952 thesis speculation regarding a Schwinger-type Lagrangian action operator, I consider the full-scale effort in 1953 by Bergmann and Schiller (BS53) \cite{Bergmann:1953aa} to formulate a Lagrangian quantum action principle, attempting to extend the Schwinger action principle to Einstein's generally covariant theory.  This same paper also looks closely not only at the Poisson bracket generators of canonical transformations that correspond to infinitesimal coordinate transformations (invariant transformations), but also transformations between physically distinct solutions of Einstein's equations, an analysis I take up in Section 4, where I also consider the related 1955 work by Bergmann and I. Goldberg (BG55)  \cite{Bergmann:1955ag}. They undertook a closer examination of the classical factor group of canonical transformations modulo invariant transformations, leading to the notion of a reduced phase space whose elements are constants of the motion.  The algebra of these observables was realized by what I call `Bergmann-Goldberg brackets' - modestly termed `extended Dirac brackets' by the authors.  With the recognition that the determination of these brackets required the construction of diffeomorphism invariants (constants both temporally and spatially), I consider in Section 5 efforts to construct diffeomorphism invariants. Bergmann's student Newman proposed an iterative construction. Alternatively, Janis investigated the possibility of obtaining the desired algebra of invariants through the imposition of coordinate conditions, simplified through the addition of a divergence to the Einstein Lagrangian that trivialized the primary constraints. Dirac somewhat later in 1959 (D59) \cite{Dirac:1959aa} proposed his own general relativistic coordinate conditions, with Bergmann's enthusiastic support. Roughly simultaneously Komar proposed the use of intrinsic coordinates to construct the elements of this physical phase space (K58) \cite{Komar:1958aa}. 
 
  In Section 6 I discuss Dirac's 1958 gravitational Hamiltonian (D58) \cite{Dirac:1958aa} and Bergmann's 1962 (B62) \cite{Bergmann:1962ac} interpretation. This reformulation of general relativistic Hamiltonian dynamics strongly influenced Bergmann's own treatment of general covariance.  Most significantly, he showed that Dirac's new variables implied that infinitesimal diffeomorphisms must be understood as incorporating a compulsory metric field dependence - that which is now taken as a conventional decomposition in the direction normal to the spacelike temporal foliations of spacetime. This decomposition later became a foundational principle in Wheeler's geometrodynamics program - but with a conscious abandonment of the full four-dimensional symmetry that had been pursued by the Syracuse group. I will argue that although the Wheeler program has until recently dominated the efforts at quantization, the Bergmann program has itself undergone a continuous development since the supposed pre-Renaissance era, and is in my opinion now poised to reassert what Pitts in this volume has called its strong general relativity exceptionalism. Significantly, the lapse and shift were abandoned as phase space variables. The link to the underlying four-dimensional diffeomorphism symmetry was thus obscured, and perhaps even temporally lost. Such was the case in the equivalent parallel work of Arnowitt, Deser, and Misner, culminating in their 1962 review \cite{Arnowitt:1962ab}. Bergmann carefully laid out the basis of this apparent loss in his 1962 interpretation of Dirac's classical general relativistic Hamiltonian analysis \cite{Bergmann:1962ac}.
 
 \section{Initial consequences of a generally covariant Lagrangian action}
 \label{two}
Given that a Lagrangian density transforms under an arbitrary infinitesimal spacetime coordinate transformation $x'^\mu = x^\mu + \xi^\mu(x)$ as a density of weight one plus a possible total divergence, Bergmann showed in 1949 \cite{Bergmann:1949aa}  that the Legendre matrix must be singular and that consequently there exists as many (primary) constraining relations among the fields and conjugate momenta as there are independent null vectors of this matrix. Bergmann and Brunings \cite{Bergmann:1949ab} then applied this logic to a generic reparameterization covariant theory with field variables $y_A(x)$ and with spacetime coordinates $x^\mu$ themselves functions of four parameters $u^\alpha :=(t, u^s)$. They assumed a Lagrangian density 
$\mf{L} $
 that depended only on field variables $y_A$ and first derivatives with respect to the spacetime coordinates, $y_{A,\mu} := \frac{\partial y_A}{\partial x^\mu}$. BPSZ50 \cite{Bergmann:1950ab}  later assumed that the  Lagrangian was quadratic in the $y_{A,\mu}$, and of the form 
 \beq
 \mf{L} = \Lambda^{A \rho B \sigma}(y) y_{A,\mu} y_{B, \nu}, \label{quadL}
 \eeq
 then the transformed Lagrangian with variables $x^\mu(u)$ is no longer  quadratic in time derivatives $y_{A,0}$ since $y_{A,\mu} =\frac{\partial u^\alpha}{\partial x^\mu} \frac{\partial y_A}{\partial u^\alpha} = J^{-1} M^\alpha{}_\mu  \frac{\partial y_A}{\partial u^\alpha}$, where the $M^\alpha{}_\mu $ are minors of the matrix $\frac{\partial x^\nu}{\partial u^\beta}$\footnote{Explicitly, $M^\alpha{}_\mu = \frac{1}{3!}\epsilon^{\alpha \beta \gamma \delta}\epsilon_{\mu \nu \rho \sigma}\frac{\partial x^\nu}{\partial u^\beta} \frac{\partial x^\rho}{\partial u^\gamma}\frac{\partial x^\sigma}{\partial u^\delta}$} and $J$ is the determinant of this matrix. Both $J$ and $M^\alpha{}_\mu $ are functions of $\dot x^\mu := \frac{\partial x^\mu}{\partial t}$. It does turn out, however, that the resulting Lagrangian is homogeneous of degree one in the velocities $\dot y_A$ and $\dot x^\mu$.
They simplified the notation by representing the variables $y_A(t,u^s)$ and $x^\mu(t, u^s)$ collectively as $y_a$, with corresponding conjugate momenta $\pi^a$.  Then the Hamiltonian density defined in the usual manner as $\mf{H} :=  \pi^a \dot y_a - \mf{L} \equiv 0$ vanished identically when viewed as a functional of the fields and velocities. Seven of the primary constraints could easily be found. The challenge was to find the eighth which they showed needed to be employed as the Hamiltonian density (with the option of adding arbitrary linear combinations of the remaining seven). They were able to demonstrate that the original Lagrangian dynamics followed from the resulting Hamiltonian field equations. 

The expressed rationale for the introduction of the parameter formalism was that in this way if would be possible to incorporate singular particle sources in the theory. The parameterization procedure was apparently inspired by the work of Paul Weiss who had employed it in work cited by Bergmann and Brunings \cite{Weiss:1936aa,Weiss:1938aa,Weiss:1938ab}.\footnote{Goldberg has expressed to me the opinion that the parameterization idea was inspired by Weiss.  Weiss had introduced a parameterized version of electromagnetism in flat spacetime, but he did not conceive of the $x^\mu(u)$ as dynamical variables. See \cite{Rickles:2015aa} for more background, and also \cite{Salisbury:2020ac} for Weiss' influence on Hamilton-Jacobi techniques in general relativity.} It is possible that at least initially Bergmann thought it conceivable that just as the classical field theory automatically determined particle motions via the Einstein Infeld Hoffmann (EIH) procedure, the quantized field might automatically induce particle quantum wave mechanics.\footnote{See  \cite{Blum:2018aa}. See also \cite{Salisbury:2020ac} and \cite{Salisbury:2020ab} for a discussion of the Syracuse school's work on classical equations of motion.} He did write that ``it is possible, in the general theory of relativity, to treat the motion of field singularities (which are used to represent particles) without having to deal with infinite interaction terms of one kind or another. It is possible that this accomplishment will lead to a more satisfactory quantized theory \ldots "\footnote{\cite{Bergmann:1949aa}, p. 680}

The task of constructing the explicit Hamiltonian was taken up in BPSZ50 \cite{Bergmann:1950ab}. Since the Legendre matrix $G^{ab}:=\frac{\partial^2 \mf{L}}{\partial \dot y_a \partial \dot y_b}$ is singular, it is not possible to solve the defining equation $\pi^a = G^{ab} \dot y_b + f^a[y]$ for the velocities in terms of the fields $y_a$ and momenta $\pi^b$. BPSZ50 developed a scheme for solving for the $\dot y_b$ which involved the use of a matrix $E_{ab}$ that they called a `quasi-inverse' of $G^{ab}$ which satisfied the relations $G^{ab} E_{bc} G^{cd} = G^{ad}$ and $E_{ab}G^{bc} E_{cd} = E_{ad}$.\footnote{Equations (3.1) in BPSZ50 have incorrect right hand sides. The typos were corrected in \cite{Anderson:1951aa}, equation (4.3). }  They did not cite a reference for this procedure.\footnote{The matrix $E_{ab}$ is known as the Moore-Penrose pseudoinverse \cite{Rao:1971aa}, first described by Moore in 1920 \cite{Moore:1920aa}, and then rediscovered by Penrose in 1955 \cite{Penrose:1955aa}. Penrose has communicated to me that he was unaware of its use in BPSZ50.} The bulk of this paper dealt with the determination of $E_{ab}$. Since the task was greatly simplified when the parameterized formalism was abandoned, I will wait to describe this simplified procedure. Most relevant for the moment is the fact that the resulting vanishing Hamiltonian density was quadratic in the momenta, taking the form $\mf{H} = \frac{1}{4} \pi^a E_{ab} \pi^b = 0$. Taking as the gravitational Lagrangian Einstein's first-order form 
\beq
\mf{L}_E := \sqrt{-g}g^{\mu \nu} \left(\Gamma^\sigma_{\rho \sigma} \Gamma^\rho_{\mu \nu} -\Gamma^\sigma_{\rho \mu} \Gamma^\rho_{\sigma \nu} \right) \label{EL},
\eeq
they found an explicit expression of the Hamiltonian that generated the correct first-order equations of motion for all of the phase space variables. As expected, it contained eight arbitrary spacetime functions, four of them corresponding to the original diffeomorphism freedom, and four to the reparameterization freedom. The first four  were functions that appeared in the solution for $E_{ab}$. The fifth amounted to an arbitrary choice for the function $t(x)$, and the three remaining functions could be chosen to multiply the constraints that generated reparamterizations in $u^s$. Still in this work the authors stated their intention that `the new formalism will be used to give a new derivation of the equations of motion.'\footnote{\cite{Bergmann:1950ab}, p. 88. Curiously, even though Joshua Goldberg had just begun his own thesis work \cite{Goldberg:1952ab} under Bergmann's direction at this time, and it was devoted both to the covariance foundations of the EIH approach and the Hamiltonian form of general relativity, he has communicated to me that he never shared this view.}

At this juncture we make a quick detour to the Hamiltonian derivation of Pirani and Schild \cite{Pirani:1950aa}. They had listened to Dirac's Vancouver lectures on constrained Hamiltonian dynamics in August of 1949. A letter from Bergmann to Schild, dated February 12, 1949, indicates that Schild had already read  Bergmann's 1949 paper.\footnote{``I am sorry that we did not get together in New York \ldots In the meantime, I have received the reprints you were kind enough to send me. I shall reciprocate in kind and send you one I have here at the moment and regularly send you my output \ldots We are having a project at Syracuse in which we are attempting to investigate the general properties of covariant field theories (whether or not they be impressed on a Riemannian geometry), with the special purpose of learning how to quantize covariant theories. If you should have any papers relating to this general subject, I should certainly appreciate knowing about them."} It was natural that it would then have occurred to him that Dirac's procedure could be applied to general relativity, and he suggested this as a thesis topic to Pirani. Even though they continued to work with the Bergmann Brunings parameterized model, the technique they employed turned out to be substantially less involved than BPSZ50. 
Dirac's procedure involved the independent variation of $y_A$, $\pi^B$, and $\dot y_A$, leading him to distinguish a `strong' equality which remained valid under these variations, assuming one begins on the constraint hypersurface in phase space, and a `weak' equality which is not preserved under these variations. He could prove that with constraints $\phi_a = 0$ the Hamiltonian could be written as a strong equality, $H \equiv \beta_a \phi_a$, where the $ \beta_a$ are arbitrary functions of the three sets of variables. Some ingenuity was still required to construct the Hamiltonian constraint which was assumed to be of the form $H = \dot x^\gamma \left(\Phi_\gamma + c_{\gamma \sigma} \phi^\sigma\right)$ where the $ \phi^\sigma$ were already known constraints. The problem was to find the constraints $\Phi_\gamma$. This feat was accomplished through the adroit addition of a strongly vanishing product of known constraints. Goldberg insists that the news in Syracuse of the Pirani Schild result was a shock and the Dirac analysis was still unknown when the preprint arrived.\footnote{private communication.} The resulting exchange was nevertheless civil. Bergmann acknowledged receipt of the preprint in a letter to Schild dated January 15, 1950: ``I got your letter, plus preprints, in Syracuse last week, in the midst of reading term papers. As a result, I have not been able to compare your formula with ours to see whether they agree. This is, however,  trivial, because disagreement would mean an arithmetical error on our part or yours, and that is not very likely. The formulas will look quite different, and that is why comparison will be somewhat slow."  He continued ``From the external appearance of your constraint, I suspect that you get yours by methods quite different from ours, and thus, I think we ought to publish separately, calling attention to the other fellow's paper. In particular, we are distributing a preprint of our paper as an ONR Technical Report (as soon as it is off the press, you will get your copy). I would, therefore, suggest that you people in Pittsburgh go ahead with your plans for publication, and we do with ours... From the point of view of further development, I consider it extremely important that we should exchange complete information on the technique employed, because the obvious next stage is to construct  improved covariant theories, and to do so requires use of the tricks learned. Likewise, we should know what we are planning, so that we shall not duplicate our work unnecessarily. So I am looking forward to spending a day with you in New York. Give me a ring as soon as you hit town.'"The reference is to the APS meeting in New York, February 2 - 4, 1950 in which either Schild or Pirani reported on their work.

Bergmann had already by this time had some misgivings about the BPSZ50 results, as reported by Scheidegger to Schild on January 20, 1950.\footnote{See also \cite{Blum:2018aa}} Scheidegger writes ``I had a chance to talk to Bergmann; he is still worried about the quantization of the gravitational field; he thinks that his previous results were wrong since there are some 8 more constraints which had been overlooked heretofore." Indeed, as I have noted previously \cite{Salisbury:2012aa} , Bergmann and Brunings had stated without supporting evidence that there were no constraints beyond the eight primary ones. A letter from Bergmann to Schild dated November 16, 1950 confirms his realization that more constraints are present: ``In the meantime we have continued our work and found that additional constraints must be satisfied if the equations of the Hamiltonian formalism are to be equivalent to those of the Lagrangian formalism. At the present we are working on a proof that the number of these additional constraints equals the number of primary constraints. A serious quantization procedure must take all the constraints into account ab initio. As soon as we know, we shall write up and send you a copy of the MS. What are you doing these days?" Schild responds on February 5, 1951: ``Felix and I are very interested in your results on additional constraints. At the moment I do not quite understand how they come in. In Dirac's language, these constraints mean that the identities obtained earlier by you (and presumably the same as those obtained by us) are second class $\phi$ - equations. We had an argument that seemed convincing at the time, that the Poisson brackets between identities for the gravitational Lagrangian ought to vanish. Your results seem to contradict this (Do you get new constraints in the specific case of the gravitational field without singularities?) We would be very interested to see your manuscript when you have the mimeographed form. We are looking into the matter too and will let you know as soon as we can confirm your results." According to a note from  Schild to Bergmann dated January 23, 1952, Schild's student R. Skinner began working on this problem in June, 1951, as a component of his dissertation. The results were published jointly with Pirani and Schild \cite{Pirani:1952aa}. By this time they had abandoned the parameterized formalism, as had AB51 in 1951 \cite{Anderson:1951aa}.

\subsection{The Rosenfeld, Anderson, Bergmann, Penfield, Schiller nexus}

Here the story becomes confused. Goldberg maintains that Bergmann had always known that there were additional constraints. Yet he attributes the first derivation of secondary constraints in general relativity to Penfield, following Penfield's discovery that the parameterized formalism could be abandoned.\footnote{private communication, and cf. also \cite{Salisbury:2020ab}} Penfield showed that the four vacuum Einstein equations $G_{\mu \nu}= 0$ followed as a consequence of the preservation in time of the primary constraints \cite{Penfield:1950aa}. Goldberg went on to show in his thesis that these secondary constraints were preserved as a consequence of the Bianchi identities \cite{Goldberg:1952ab}. Also, Penfield was listed as a co-author in the abstract of the APS talk that announced the results of the AB51 paper that has until recently been cited as the first work in which the secondary constraints of Einstein's theory were systematically derived. The tale is substantially further complicated through the discovery by the Syracuse group in 1950 of Rosenfeld's pioneering systematic analysis of the constrained Hamiltonian dynamics of Einstein's theory\cite{Rosenfeld:1930aa,Rosenfeld:2017ab}. Anderson informed me in 2007 that it was he who had made this discovery and brought it to Bergmann's attention. In the same conversation Schiller told me that this Rosenfeld paper inspired his own doctoral thesis \cite{Schiller:1952aa}, and he cited Rosenfeld both in the original thesis and is cited in this article which was received by the Physical Review in May, 1951. They write that their work ``is in some respects similar to results obtained by Rosenfeld."  This is of particular relevance since the same methods that Rosenfeld had developed for a generally covariant Lagrangian theory were directly applicable to the quadratic model with the Einstein Lagrangian (\ref{EL}) in its unparameterized form, as analyzed by Anderson and Bergmann \cite{Anderson:1951aa}. Furthermore, Rosenfeld had already in this groundbreaking article not only developed an algorithm for determining all higher order constraints, but also showed how to construct the phase space generator of general infinitesimal coordinate transformations.\footnote{See \cite{Salisbury:2017aa} for a detailed analysis of this paper.} Penfield's thesis, which he defended in July, 1950, contains no reference to Rosenfeld. The published version \cite{Penfield:1951aa} is almost identical - excepting for the inclusion of this reference, and this - in addition to Penfield's full time teaching duties at Harpur College - may account for the one year delay in its submission to the Physical Review. Bergmann's first public recognition of Rosenfeld's work apparently occurred at the August 1950 International Congress of Mathematicians \cite{Bergmann:1952aa}.

The basis of all of the cited works is the fundamental identity that follows from the assumption that the variation under the symmetry group of the Lagrangian density is a total divergence. Although all of the authors considered more general symmetry groups, I will confine my attention here to general covariance under infinitesimal coordinate transformations $x'^\mu = x^\mu + \xi^\mu(x)$, in which case all assumed that the fields $y_A$ transformed as
\beq
\bar \delta y_A = F_{A \mu}{}^{B\nu} y_B \xi^\mu_{, \nu} - y_{A,\mu} \xi^\mu, \label{deltay}
\eeq
where the $F_{A \mu}{}^{\nu \sigma}$ are constants. Furthermore, it was assumed that the term $S^\mu_{,\mu}$ that was added to the Lagrangian was linear in the second derivatives, so that $S^\mu = f^{A \mu \rho}(y) y_{A,\rho}$. 
In this case the sum of the Lagrangian density and a total divergence must transform as a scalar density of weight one, namely\footnote{This is a direct expression of the transformation property of a scalar density $\mf{S}\left(y(x)\right)$ of weight one under a coordinate transformation $x'^\mu(x)$, $\mf{S}'\left(y'(x')\right) = \mf{S}\left(y(x)\right) \left|\frac{\partial x}{\partial x'}\right|$}
\beq
\bar \delta \mf{L} + \left(\bar \delta S^\mu \right)_{, \mu} \equiv - \left(\left(\mf{L}+ S^\nu_{,\nu}\right)\xi^\mu\right)_{,\mu}. \label{bardelta}
\eeq
This is true in the passage from $\sqrt{-g}R$ to the Einstein Lagrangian, as Bergmann had spelled out in detail in his textbook \cite{Bergmann:1942aa}.  Although Rosenfeld did not employ the Einstein Lagrangian, it was quadratic in the first derivatives of the field variables. The difference was that his purely gravitational contribution was constructed using tetrad fields, and as such it transformed as a scalar density under general coordinate transformations  whereas the Einstein Lagrangian did not.  On the other hand under local Lorentz transformations its variation involved a total divergence, and Rosenfeld showed how to construct the correct Hamiltonian with this additional local gauge symmetry. 

Since (\ref{bardelta}) is an identity, a wealth of information can be derived from it. It must be true that the coefficients of each derivative of $\xi^\mu$ must vanish. The origins of this procedure can be traced all the way back to Felix Klein \cite{Klein:1918aa}.\footnote{See \cite{Salisbury:2017aa}} This is the method utilized by Rosenfeld. Bergmann, on the other hand, undertook an integration over all space, taking the $\xi^\mu \rightarrow 0$ on the boundary. In this manner he derived what he called `the generalized contracted Bianchi identities,'
\beq
F_{A \mu}{}^{B \rho} \left(L^A y_B \right)_{,\rho} + L^A y_{A,\mu} \equiv 0, \label{nullv}
\eeq
where $L^A := \frac{\partial \mf{L}}{\partial y_A} - \left(\frac{\partial \mf{L}}{\partial y_{A, \mu}}\right)_{, \mu}$. It follows that the coefficient of $y_{A,\mu \nu \rho}$ must vanish identically, leading to the conclusion that $F_{A \mu}{}^{B 0} y_B \Lambda^{AC} \equiv 0$. Using Bergmann's notation I let (\ref{quadL})  represent  the Einstein Lagrangian (\ref{EL}). Then it follows from (\ref{nullv}) that $u_{\mu A} := F_{A \mu}{}^{B 0} y_B$ is a null vector of  $\Lambda^{AC}:= \Lambda^{A0B0}$,\footnote{This two-index object should not be confused with the parameters introduced in BPSZ50.}
\beq
u_{\mu A} \Lambda^{AC} \equiv 0.
\eeq
Then since 
\beq
\pi^A = 2 \Lambda^{A0B0} \dot y_B +2  \Lambda^{A0Ba} y_{B,a} \label{piA}
\eeq
 it follows that (\ref{piA}) does not possess unique solutions for $\dot y_A$. Rosenfeld's procedure for obtaining the general solution was to first find those linear combinations of $\dot y_A$ which could be set equal to zero, consistently with (\ref{piA}). He let $\dot y^0_A$ represent this solution set, including the non-vanishing $\dot y_A$. For his Lagrangian choices there was a simple procedure for finding this special solution. Letting $N$ represent the number of fields, he merely needed as a first step to isolate $N-4$ independent rows of the matrix $\Lambda$. Then he could quickly find the four null vectors $u_{s A}$, where $s = 1 , \ldots, 4$. Furthermore, $N-4$ of the equations of (\ref{piA}) could be solved for $N-4$ momenta, and the remaining momenta could be expressed in terms of them. And finally, the general solution of (\ref{piA}) is $\dot y_A = \dot y^0_A + \lambda^s u_{s A}$ where the $ \lambda^s(x)$ are arbitrary spacetime functions.\footnote{See \cite{Salisbury:2017aa} for details} When this solution is substituted into the lagrangian the null vectors do not contribute so that the Lagrangian then becomes a function of $y_A$ and $\pi^B$. On the other hand the contribution to the Hamiltonian from $\pi^A \dot y_A$ becomes a function of $y_A$ and $\pi^B$, plus a sum of primary constraints multiplying the arbitrary functions $\lambda^s$. Although Rosenfeld for some unknown reason did not calculate the full Hamiltonian for his tetrad field, he could easily have done so using his methods, as shown in \cite{Salisbury:2017aa}.
 
I have gone through Rosenfeld's method in some detail to compare and contrast with the Bergmann school procedure. Robert Penfield was actually the first to obtain the unparameterized gravitational Hamiltonian, and he did so by using the quasi-inverse $E_{AB}$ which satisfies the relations 
\beq
\Lambda^{AB} E_{BC} \Lambda^{CD} = \Lambda^{AD}, \label{G}
\eeq
 and 
 \beq
 E_{AB}\Lambda^{BC} E_{CD} = E_{AD}. \label{E}
 \eeq
  His thesis \cite{Penfield:1950aa} contains slightly more detail than the published work \cite{Penfield:1951aa}. The basic idea is to use the matrix $D^C{}_A$ that transforms the matrix $\Lambda^{AB}$ into a `bordered' matrix in which the final four rows and columns vanish. This feat is accomplished using the four null vectors $u_{s A}$ as the final four rows, i.e., $D^{N-4+s}{}_A = u_{s A}$, making certain that the first $N-4$ rows are linearly independent. Using AB51's notation, I will represent the bordered matrix with a prime, so that therefore $\Lambda'^{CD} =  \Lambda^{AB} D^C{}_A D^D{}_B$. Continuing with their notation I represent the resulting $N-4$ dimensional non-vanishing invertible matrix by $\Lambda'^{C^*D^*} $, where the starred indices range from $1$ to $N-4$. Let the inverse be $G_{A^*B^*}$. 
It results that $E_{AB} = G_{C^*D^*} D^{C^*}{}_A D^{D^*}{}_B$. 

Penfield did not refer explicitly to this quasi-inverse. Rather, he noted that corresponding to the bordering procedure, there was a related transformation of the $\dot y_A$, namely $\dot y'_A = D^{-1}{}^C{}_A \dot y_C$ such that $\Lambda^{AB} \dot y_A \dot y_B = \Lambda'^{AB}  \dot y'_A \dot y'_B  = \Lambda'^{A^*B^*}  \dot y'_{A^*} \dot y'_{B^*}$. Therefore only these specific linear combinations of the velocities were fixed by (\ref{piA}).  Furthermore, (\ref{piA}) can be solved for these velocities as follows. First contract with $D^C{}_A$ to get
\beq
D^C{}_A \pi^A = 2 \Lambda'^{C D} \dot y'_D + 2 D^C{}_A \Lambda^{A B a} y_{B,a}, \label{pip}
\eeq
yielding  $D^{C^*}{}_A \pi^A = 2 \Lambda'^{C^* D^*} \dot y'_{D^*}+ 2 D^{C^*}{}_A \Lambda^{A B a} y_{B,a}$, which can be solved to give 
\beq
 \dot y'_{A^*} = \frac{1}{2}G_{A^* C^*}\left(D^{C^*}{}_B \pi^B - 2 D^{C^*}{}_D \Lambda^{D B a} y_{B,a}\right)
 =:  \frac{1}{2}G_{A^* C^*} D^{C^*}{}_D \bar \pi^{D}. \label{ydotp}
 \eeq
 Note that therefore velocities $\dot y^0_A$ that can be expressed in terms of the momenta are
 \beq
 \dot y^0_A = \frac{1}{2} D^{B^*}{}_A D^{C^*}{}_D G_{B^* C^*} \bar \pi^D,
 \eeq
 and the general solution is
 \beq
 \dot y_A = \dot y^0_A  + u_{\alpha A} w_\alpha = \frac{1}{2} D^{B^*}{}_A D^{C^*}{}_D G_{B^* C^*} \bar \pi^D+ u_{\alpha A} w_\alpha. \label{ydot}
 \eeq
  In addition, taking $C = N-\alpha$ in (\ref{pip}), we get the primary constraints
\beq
g_\alpha := D^{N-\alpha}{}_A\left( \pi^A -  2  \Lambda^{A B a} y_{B,a}\right) = u_{\alpha A} \bar \pi^A = 0.
\eeq
Now substituting (\ref{ydot}) observe that the Lagrangian density becomes
\bea
\mf{L}& =& \Lambda^{AB} \dot y^0_A \dot y^0_B+ 2 \Lambda^{A Ba}\left( \dot y^0_A +u_{\alpha A} w_\alpha \right)y_{B,a} + \Lambda^{A a B b}y_{A,a} y_{B,b} \nonumber \\
&=& \frac{1}{4} G_{A^* B^*}D^{A^*}{}_C D^{B^*}{}_D\bar \pi^{C}\bar \pi^{D} +  \Lambda^{A B a}D^{C^*}{}_A  y_{B,a}G_{C^*D^* } D^{D^*}{}_E\bar \pi^{E} \nonumber \\
&+&2 \Lambda^{A Ba} u_{\alpha A} w_\alpha y_{B,a} + \Lambda^{A a B b}y_{A,a} y_{B,b}.
\eea
Also
\beq
\pi^A \dot y_A =  \left(\bar \pi^A + 2 \Lambda^{A B a}y_{B,a}\right)\left( \frac{1}{2} D^{C^*}{}_A G_{C^*D^*}D^{D^*}{}_E \bar \pi^E +u_{\alpha A} w_\alpha \right)
\eeq
Therefore the Hamilton density is
\beq
\mf{H} = \pi^A \dot y_A  -\mf{L}  =   \frac{1}{4} G_{A^* B^*}D^{A^*}{}_C D^{B^*}{}_D\bar \pi^{C}\bar \pi^{D}  - \Lambda^{A a B b}y_{A,a} y_{B,b} + w^\alpha g_\alpha. \label{hamd}
\eeq
This is AB51's equation (4.9), although they did not give the explicit form for the quasi-inverse. It appears in Penfield's thesis \cite{Penfield:1950aa} and also in \cite{Penfield:1951aa}. The significance of this result cannot be overemphasized. It applies to every general relativistic Lagrangian - including the Dirac and ADM Lagrangians which differ from the Einstein Lagrangian merely in the choice of $S^\mu$. In each case we have a sum of contributions of arbitrary spacetime functions multiplying the primary constraints. 

It is noteworthy also that the procedure is essentially the same that had been developed by Rosenfeld in 1930. The velocities $\dot y'_A$ are recognized as his $\dot y^0_A$ where he chooses $\dot y'_{N-s} = 0$. Then the general solution for the velocities is obtained by adding the $w_s(x)$ multiplying the null vectors of the Legendre matrix. I maintain that if Rosenfeld had felt so inclined,  he could have undertaken a straightforward modification of his case two to address Penfield's problem. He did not because he had a much more ambitious goal: to devise a quantum theory for all known particles experiencing all known forces, including gravity. This interaction for spinorial fields required the use of tetrads.\footnote{See \cite{Salisbury:2009ab} for more context - and his correspondence  with Dirac!} It is also remarkable, as AB51 observe in a footnote, that the same construction of a particular solution $\dot y^0_A(\pi, y_B, y_{A,a})$ can be employed for the non-quadratic parameterized theory.\footnote{It is interesting that this statement applies also to the relativistic string, but this is not the approach that I took in the thesis \cite{Salisbury:1977aa} that I wrote under Bergmann's direction, nor in the follow-up preprint  \cite{Salisbury:1981aa} and publication \cite{Salisbury:1984aa}.} They use the same notation as Rosenfeld to represent it. And then as Rosenfeld noted, the general solution becomes  $\dot y_A =  \dot y^0_A + w^\alpha u_{\alpha A}$.

The main focus of AB51 is the construction of the phase space generators of infinitesimal general coordinate transformations, with the field variables transforming according to (\ref{deltay}). They begin by considering the corresponding variation of the momenta (\ref{piA}), with the remarkable conclusion that these variations do not depend on higher time derivatives of the descriptors $\xi^\mu$ that appear in (\ref{deltay}) ; Isolating the terms containing the second time derivatives we have 
\beq
\bar \delta \dot y_A = F_{A \mu}{}^{B 0} \ddot \xi^\mu + \ldots = u_{\mu A}  \ddot \xi^\mu + \ldots \label{deltaydot}
\eeq
 But according to (\ref{piA}), $\bar \delta \pi^A = 2 \Lambda^{AB} \bar \delta \dot y_B + \ldots$ and there is therefore no contribution from $\ddot \xi^\mu$ since $\Lambda^{AB}u_{\mu B} = 0$.   This presents a puzzle that only much later did Bergmann identify in print, ``During the early Fifties those of us interested in a Hamiltonian formulation of general relativity were frustrated by a recognition that no possible canonical transformations of the field variables could mirror four dimensional coordinate transformations and their commutators, not even at the infinitesimal level. That is because (infinitesimal or finite) canonical transformations deal with dynamical variables on a three-dimensional hypersurface, a Cauchy surface, and the commutator of two such infinitesimal transformations must be an infinitesimal transformation of the same kind. However, the commutator of two infinitesimal diffeomorphism involves not only the data on a three-dimensional hypersurface but their `time'-derivatives as well. And if these data be added to those drawn on initially, then, in order to obtain first-order `time' derivatives of the commutator, one requires second-order `time' derivatives of the two commutating diffeomorphisms, and so forth. The Lie algebra simply will not close."\footnote{\cite{Bergmann:1979aa}, p. 175}   In more detail, suppose one carries out an infinitesimal transformation $x^\mu_1= x^\mu + \xi_1^\mu(x)$, followed by a second, $x^\mu(x_1) + \xi^\mu_2(x_1) \approx x^\mu +\xi_1^\nu(x)+  \xi^\mu_2(x)+ \xi^\mu_{2,\nu}(x)\xi^\nu_1(x) $. Then one finds the difference when the operations are carried out in reverse order. This is the Lie algebra commutator $\xi_{1,\nu}^\mu \xi_2^\nu - \xi_{2,\nu}^\mu \xi_1^\nu$.  
  But then the commutator of the infinitesimal transformation with this descriptor, commuted with a third transformation, should in principle yield second time derivatives of the original descriptors $\xi_1^\mu$ and $\xi_2^\mu$, etc. There was a clear conflict with the diffeomorphism Lie algebra since the Poisson bracket commutator of two of these generators could not yield second time derivatives of the descriptors $\xi_1^\nu$ and $\xi_2^\nu$, etc.  Nevertheless, AB51's conclusion at this time was that the symmetry generator could be expanded as 
\beq
\mf{C} = {}^0A_\mu \xi^\mu + {}^1A_\mu \dot \xi^\mu, \label{Cgen}
\eeq
  with the understanding that $\bar \delta y_A = \left\{\int d^3x \mf{C}, y_A\right\}$ and $\bar \delta \pi^A = \left\{\int d^3x \mf{C}, \pi^A\right\}$.  It did not depend on higher time derivatives of the descriptors $\xi^\mu$.
   (We will see later how this apparent inconsistency was resolved.)

AB51 then required that under an infinitesimal coordinate transformation the Hamiltonian $H = \int d^3x \mf{H} $ must retain (\ref{hamd}) in the same form. Considering the special case in which the arbitrary functions $w^\alpha$ depend explicitly on the spacetime coordinates but not on the canonical variables, this is the condition that $\bar \delta H = \int d^3x \delta w^\alpha g_\alpha$. Writing out the variation of $H$ generated by (\ref{Cgen}) we conclude that
\bea
&& \int d^3x \delta w^\alpha g_\alpha = \int d^3x\left( \left\{ \mf{C}, H \right\} +  \frac{\partial \mf{C}}{\partial t} \right) \label{deltaH1} \\
 =&& \int d^3x\left( \left\{ {}^0A_\mu, H \right\} \xi^\mu +\left\{ {}^1A_\mu , H \right\}\dot \xi^\mu +  {}^0A_\mu \dot \xi^\mu + {}^1A_\mu \ddot \xi^\mu \right), \label{deltaH2}
\eea
 where (\ref{Cgen}) was used in the second line. Next they used the fact that at the fixed time at which these spatial integrals are calculated each order of time derivative can be changed arbitrarily, and therefore the coefficients of each order of time derivative in $\delta w^\alpha$ needed to match with the corresponding coefficient on the right hand side.  First they focused attention on the arbitrary $\ddot \xi^\mu$ term in (\ref{deltaH2}). On the one hand they knew from (\ref{deltaydot}) that the only term in $\bar \delta \dot y_A$ that is dependent on $\ddot \xi^\mu$ is $u_{\mu A} \ddot \xi^\mu$. But from (\ref{ydot}) they knew that the $\ddot \xi^\mu$-dependent term in $\bar \delta \dot y_A$ must be $\delta w^\mu u_{\mu A}$ since the variation of the phase space variables does not depend on this higher time derivative. It follows that $\delta w^\mu = \ldots + \ddot \xi^\mu$. So finally they concluded that the ${}^1A_\mu$ must be the primary constraints $g_\mu$. 

 Next, regarding the coefficient of $\dot \xi^\alpha$ on the right hand side of (\ref{deltaH2}) it is clear from the left hand side  of (\ref{deltaH1}) that it must be proportional to the primary constraints, i.e.,
$\left\{ {}^1A_\mu , H \right\} +  {}^0A_\mu$ must be a linear combination of the $g_\alpha$. Similarly, looking at the coefficient of $\xi^\alpha$ it follows that $\left\{ {}^0A_\mu , H \right\}$ must also be a linear combination of the $g_\alpha$. It is noteworthy that the existence of secondary constraints does not follow from these results. Rather, AB51 were aware that the time rate of change of the primary constraints ${}^1A_\mu$ needed to be set equal to zero in order to maintain consistency with the Lagrangian theory. Indeed, they wrote that if the Poisson brackets of the primary constraints with the Hamiltonian did not vanish identically, then ``they must be set equal to zero, and the requirement then becomes that the Poisson bracket of these expressions with the Hamiltonian vanish, and so on until a point is reached where no new constraints are being obtained."\footnote{\cite{Anderson:1951aa}, p. 1023}.   AB51 thus insisted that the ${}^0A_\mu$ and ${}^1A_\mu$, ``together with the hamiltonian, form a function group."\footnote{\cite{Anderson:1951aa}, p. 1023} It is noteworthy that they did not deduce that the Hamiltonian itself must be a linear combination of constraints, i.e., that it must vanish. 

AB51 have until recently been identified as having been the first to demonstrate that secondary constraints arise in general relativity. But that honor really belongs to Rosenfeld.  He actually proved it in an argument of the type that Bergmann and Schiller later employed in 1953. His starting point was the identity (\ref{bardelta}) - although as I pointed out earlier he really looked at two special cases that did not include the Einstein Lagrangian.\footnote{See \cite{Salisbury:2017aa}, and in particular equation (37).} One immediately deduces from the identity (\ref{bardelta}) the existence of a (vanishing) conserved charge, noting that $\bar \delta \mf{L} = \mf{L}^A \bar \delta y_A + \left(\frac{\partial \mf{L}}{\partial y_{A,\mu}}\right)_{,\mu} $ and rewriting (\ref{bardelta}) as
\beq
0 \equiv \mf{L}^A \bar \delta y_A + \left(\frac{\partial \mf{L}}{\partial y_{A,\mu}}\bar \delta y_A + \bar \delta S^\mu - \left(\mf{L}+ S^\nu_{,\nu}\right)\xi^\mu\right)_{,\mu}.
\eeq
So when the field equations are satisfied, i.e. $\mf{L}^A = 0$, the current $\mf{C}^\mu :=  \frac{\partial \mf{L}}{\partial y_{A,\mu}}\bar \delta y_A + \bar \delta S^\mu - \left(\mf{L}+ S^\nu_{,\nu}\right)\xi^\mu$ is conserved, $\mf{C}^\mu_{,\mu} = 0$. But since this current depends on the arbitrary time-dependent $\xi^\mu$ it must vanish. And furthermore, in calculating the time rate of change of $d^3 x \mf{C}$ where $\mf{C} := \mf{C}^0$, one obtains precisely the AB51 relation on the right-hand side of 
(\ref{Cgen}) where one already knows that that ${}^0A_\mu$ and ${}^1A_\mu$ are constraints. In addition, Rosenfeld showed explicitly that $d^3 x \mf{C}$ generated the correct variations of {the phase space variables under infinitesimal diffeomorphisms. It is remarkable that he also did not recognize that the vacuum general relativistic Hamiltonian vanished - even though the particular generator $d^3 x \mf{C}$ with $\xi^\mu = \delta^\mu_0$ was ostensibly the Hamiltonian - except for a possible spatial surface integral!\footnote{Josh Goldberg has informed me that Bergmann himself took some time to come to this realization.} Schiller himself used this vanishing conserved charge in his thesis \cite{Schiller:1952aa}, and he was therefore apparently the first in the Syracuse group to note its role as the generator of coordinate symmetry transformations.\footnote{ I should add parenthetically that the evidence suggests that Rosenfeld was probably aware of the daunting challenge one faced in finding a constraint algebra that corresponded to the conventional diffeomorphism Lie algebra. Regarding the algebra he confined his attention to spatial diffeomorphisms whose generators did satisfy a closed Poisson bracket algebra.This discussion is in his Section 6 \cite{Rosenfeld:1930aa}\cite{Rosenfeld:2017ab}. See also \cite{Salisbury:2017aa}, pp 43-44.}

\section{Anderson, Bergmann and  Schiller, and Lagrangian approaches to quantum gravity}

After reviewing the fundamentals described above, roughly the latter half of Anderson's 1952 thesis  \cite{Anderson:1952aa} was devoted to a quantum gravitational attempt modeled on Schwinger's Lagrangian approach. The idea was that a Lagrangian approach could prove to be simpler to implement given that in configuration-velocity space the primary constraints are identities. The hope was then to deduce a quantum commutation relation amongst field variables and time derivatives by assuming, as did Schwinger, that the variation of quantum transition amplitudes was generated by an Hermitian operator. He proposed an approximation procedure for implementing this idea, assuming a quadratic Lagrangian operator of the form 
\beq
\hat L = \hat y_{A, \mu} \hat \Lambda^{A \mu B \nu}  \hat y_{B, \nu} + \hat \Delta, \label{hatL}
\eeq
 where $\hat \Lambda$ and $\hat \Delta$ are functions of the undifferentiated $\hat y_A$. He notes that in considering variations $\delta \hat y_A$ he cannot alter the order of factors, and it is therefore impossible to deduce  the field equations in the conventional manner (say with all $\delta \hat y_A$ to the right). He argues that c-number variations are excluded since then it would not be possible to undertake finite transformations $\hat y_A = \hat y_A(\hat y'_B)$. His conclusion, adopted later by Bergmann and Schiller in 1953, is that the quantum Lagrangian may not be associated with a general variational principle. He does require, instead, that the varied Lagrangian differ identically from the original only by a total divergence, and this does lead to the canonically quantized Hamiltonian  approach with the usual cast of primary and secondary constraints. 

Thus Anderson proposed to implement a Schwinger variational principle of the form
\beq
\delta \left<  \alpha'_1, t_1 | \alpha'_2 {, t_2}\right> = <  \alpha'_1, t_1 | \delta \int d^4x \hat L |\alpha'_2 , t_2>,
\eeq
where the $\alpha'$ constitute a complete set of eigenvalues of the operators $\hat \alpha$. It is noteworthy that his approximation procedure  was modeled after the Gupta-Bleuler procedure in quantum electrodynamics, in which the vanishing of the positive frequency components of gauge constraint is employed as a condition on quantum states. Anderson cited in this regard the Schwinger and Feynman techniques as expounded in Dyson's 1951 Cornell lectures \cite{Dyson:2007aa} and his groundbreaking 1949 paper \cite{Dyson:1949aa}.\footnote{Anderson explained to the author and Rickles in 2011 that his exposure to Dyson did not come from Syracuse, but rather from a visit to Mexico in the summer of 1951 where he worked with Alejandro Medina and ``tried to understand Dyson's paper, and the renormalization program. So then I got very much involved in quantum field theory."}

In 1953 Bergmann and his student Ralph Schiller BS53 \cite{Bergmann:1953aa} decided to pursue a Schwinger-type Lagrangian approach to quantum gravity.  They claimed, however, that it was not possible to formulate a Schwinger-like quantum action principle that would be valid for this wider class of solution variations. Indeed, if this had been possible then it presumably would have been possible to deduce a quantum commutator algebra for all the gravitational metric variables. Instead, they showed that in the vacuum case it was sufficient to restrict the variations to those engendered by general coordinate transformations - even though, as mentioned above, the generator of these variations vanished. Indeed, in the Lagrangian formulation the primary constraints vanish identically, and the remaining constraints are nothing other than the four Einstein equations $G_{0 \mu} = 0$ that do not involve second time derivatives of the metric. 

As in the papers cited previously, BS53 considered arbitrary generally covariant theories, but now with field operator variables $\hat y_A$ (with the `hat' signifying an operator) described by a Lagrangian $L(\hat y_A,\hat y_{A, \mu})$, and for the purpose of this discussion of the differences that arise with operators, I will spell out the new identities that arise under general coordinate transformations as a consequence of operator factor ordering. I work with Anderson's quadratic Lagrangian (\ref{hatL}). Assuming a change in coordinates $x'^\mu = x^\mu + \xi^\mu$, the corresponding variations of $y_A$ are $\bar \delta \hat y_A = F_{A \mu}{}^{B \nu}\hat y_B \xi^\mu_{, \nu} - \hat y_{A,\mu} \xi^\mu$ where the $F_{A \mu}{}^{B \nu}$ are constants and $\bar \delta \hat y_A(x) := \hat y'_A(x) - \hat y_A(x)$ is minus the Lie derivative in the $\xi^\mu$ direction. The variation in the Lagrangian operator is
$$
\bar \delta L =  \bar \delta \hat y_{A, \mu} \hat \Lambda^{A \mu B \nu}  \hat y_{B, \nu} +  \hat y_{A, \mu} \hat \Lambda^{A \mu B \nu}  \bar \delta \hat y_{B, \nu}+ \left\{\frac{ \partial\hat \Delta}{\partial \hat y_C} \cdot \bar \delta \hat y_C \right\} + \hat y_{A, \mu} \left\{\frac{ \partial\hat \Lambda^{A \mu B \nu}}{\partial \hat y_C} \cdot \bar \delta \hat y_C \right\} \hat y_{B, \nu},
$$
where the curly bracket and dot notation employed by BS53 denotes the insertion of the variation where the vacancy occurs in each derivative. Continuing, we have
$$
\bar \delta L = \left( \bar \delta \hat y_A \hat \Lambda^{A \mu B \nu}  \hat y_{B, \nu} +  \hat y_{A, \nu} \hat \Lambda^{A \nu B \mu} \bar \delta \hat y_B  \right)_{, \mu} + \left\{\frac{ \partial\hat \Delta}{\partial \hat y_C} \cdot \bar \delta \hat y_C \right\} + \hat y_{A, \mu} \left\{\frac{ \partial\hat \Lambda^{A \mu B \nu}}{\partial \hat y_C} \cdot \bar \delta \hat y_C \right\} \hat y_{B, \nu}
$$
$$
-  \bar \delta \hat y_A \left( \hat \Lambda^{A \mu B \nu}  \hat y_{B, \nu}\right)_{, \mu} - \left( \hat y_{A, \nu} \hat \Lambda^{A \nu B \mu} \right)_{,\mu}\bar \delta \hat y_B
$$
$$
:=  \left( \bar \delta \hat y_A \hat \Lambda^{A \mu B \nu}  \hat y_{B, \nu} +  \hat y_{A, \nu} \hat \Lambda^{A \nu B \mu} \bar \delta \hat y_B  \right)_{, \mu} + \left\{\hat L^A \cdot  \bar \delta \hat y_A\right\}.
$$
It is presumed that this variation satisfies identities that correspond in the classical realm to the Lagrangian varying as a density of weight one plus a total divergence, as discussed earlier. So as in the classical case one ends up with three identities, namely the vanishing of the coefficients of each order of time derivative of the arbitrary c-number descriptors $\xi^\mu$. The crucial result is that the authors determine that the following sixteen quantum operator field equations must vanish,
\beq
\left\{\hat L^A \cdot F_{A \mu}{}^{B \nu} \hat y_B\right\} = 0.
\eeq
The authors maintain, although they do not give a proof, that the classical Einstein field equations (and Einstein-Maxwell when the additional gauge symmetries are included) result in the limit as $\hbar \rightarrow 0$.\footnote{For vacuum general relativity, we read off from
$\bar \delta g_{\mu \nu} = - g_{\mu \nu , \alpha} \delta \xi^\alpha + 2 g_{\alpha (\mu} \delta \xi^\alpha_{, \nu)} =: - g_{\mu \nu , \alpha} \delta \xi^\alpha + F_{(\mu \nu) \rho}{}^{(\alpha \beta) \sigma} \delta \xi^\rho_{, \sigma}$ that 
$ F_{(\mu \nu) \rho}{}^{(\alpha \beta) \sigma} = 4 \delta^\sigma_{(\mu} \delta^{(\beta}_{\nu)} \delta^{\alpha)}_\rho$. The proposed field operator equations are then $\left\{\hat L^{\sigma \alpha} \cdot  \hat g_{\alpha \rho}\right\} = 0$.}
Specializing the Lagrangian symmetry to rigid translation in time they also obtain the quantum generator of time evolution in the Lagrangian framework which takes the form $\left\{\frac{\partial \hat L}{\partial \dot {\hat y}_A} \cdot \dot {\hat y}_A\right\} - \hat L := \hat H$ with the corresponding Schr\"odinger equation $ \hat H \Psi = i \hbar \frac{\partial \Psi}{\partial t}$. 

They also have a general expression for the quantum generator of general coordinate transformations, $\int d^3 x \left(\left\{\frac{\partial \hat L}{\partial \hat y_{A,0}}  \cdot  \bar \delta \hat y_A \right\} - \hat Q^0 \right)$ where $\hat Q^0$ arises from the divergence term in the variation of the Lagrangian.
 Using this generator it is possible to deduce commutation relations as does Schwinger in the Lorentz covariant case. There are, however, commutators that cannot be determined. Lorentz covariant quantum electrodynamics is cited as an example. Since one is working explicitly with Lagrangian expressions the momentum conjugate to $\hat A_0$ vanishes identically. (It is a primary constraint.) So the Lagrangian expression for the $U(1)$ generator is $\frac{1}{2 \pi} \int d^3 x \partial^{[0} \hat A^{s]} \xi_{, s}$, and requiring that $\delta \hat A_\rho = \xi_{, \rho}$ does not lead to a definite commutation relation involving $\hat A_0$. The authors' conclusion is that  `the commutation relations between the $\hat y_A$ and $\dot {\hat y}_A$ are not determined completely, and ``preliminary examination shows, however, that the variables whose time derivatives remain indeterminate are precisely the ones whose time derivatives are also indeterminate in the classical theory." Thus, the quantum theory must deal exclusively with observables, fields which have vanishing Poisson brackets with the generators of general coordinate transformation symmetries! 

\section{Generators of general canonical transformations and reduced phase space algebra}

One more significant innovation in the BS53 paper was the identification within the Lagrangian framework of a generally covariant system of transformations of the $y_A$ and $y_{A,\mu}$ that would correspond in phase space to general canonical transformations -  including changes that could alter the form of the field equations. The general class would however include transformations that produced physically distinct solutions to these equations. The authors identified as canonical transformations those that did not introduce higher time derivatives in the field equations. They then focused on those transformations that would in the classical realm not change the form of the field equations, and would therefore generally involve the addition of a divergence $Q^\mu_{,\mu}$ to the varied Lagrangian.  Representing those configuration-velocity transformations  that did not alter the equations of motion as $\bar \delta y_A = f_A(y_B,y_{B,\rho})$, they were able to show that even in the presence of constraints it was still possible to write these permissible variations in terms of the momenta $\pi^C$ as
$$
f_C = \frac{\partial}{\partial \pi^C}\left(\pi^B f_B - Q^4 \right).
$$
This feat was achieved by expressing the velocities $\dot y_A$ as functions of the momenta, satisfying constraints $g_i(y_A, y_{B,a}, \pi^C) = 0$ and arbitrary variables $w^i$. This rendered meaningful the derivatives with respect to $\pi^A$ of the velocity argument that appeared in  $f_A(y_B,y_{B,a}, \dot y_C)$. The chain rule for differentiation using these new variables was valid, however, only for functions $F$ satisfying the condition $\frac{\partial w^i}{\partial \dot y_A} \frac{\partial F}{\partial \pi^A} = 0$. As a consequence it turned out the generator $C:= \pi^B f_B - Q^4$ was independent of $w^i$.

Bergmann and I. Goldberg continued in 1955 (BG55) \cite{Bergmann:1955ab} this investigation of general canonical transformations and its subgroup of invariant transformations (i.e., those corresponding to diffeomorphisms and perhaps additional gauge symmetreis), but in this instance in phase space.  Their substantial achievement has gone largely unrecognized. They invented a new non-canonical phase space bracket which they modestly called an `extended Dirac bracket', but as I noted above I refer to it as the Bergmann-Goldberg bracket. This bracket, in a sense to be explained, uniquely expresses the algebra of diffeomorphism-invariant observables, but it is formulated in a general manner applicable to any Hamiltonian model possessing gauge symmetry. Remarkably, the resulting reduced phase space algebra of constants of the motion in general relativity is obtained without the imposition of coordinate conditions. Rather, the scheme requires the identification of null vectors of the symplectic form in addition to variables (not necessarily canonical) for the constraint hypersurface. Then the idea was that invariant variables could be identified, invariant in the sense that they do not change in the null directions. As we shall see below, this is equivalence to constructing diffeomorphism invariants.

The authors worked with a finite-dimensional phase space of dimension $2\mf{N}$. I will address later the extension to field theory. Specifically, let $\zeta^\mu$ be the configuration - momenta set $q_k, p_k$, with the canonical equations of motion $\dot \zeta^\mu = \epsilon^{\mu \nu} \frac{\partial H}{\partial \zeta^\nu}$, where 
$$
\epsilon^{\mu \nu} = \left(\begin{matrix} 0 & I \\ -I & 0 \end{matrix} \right).
$$
Let $C^a(\zeta^\mu ) = 0$, $a =1, \dots, N $, represent constraints.  Now confine attention to the constraint hypersurface which we assume to be covered by the phase space functions $y^m$}. It is convenient to let $Y^\alpha := y^m, C^a$ cover the full $2 \mf{N}$ dimensional phase space. BG55 showed that those transformations $\delta y^m$ on the constraint hypersurface that preserve the constraints are generated by functions $A(y)$, such that
\beq
\epsilon_{mn} \delta_A y^n = \frac{\partial A}{\partial y^m},  \label{epdelta}
\eeq
where $\epsilon_{mn} := \zeta^\mu_{,m} \epsilon_{\mu \nu} \zeta^\nu_{,n}$.

Most interesting for us at the moment is the situation in which $\epsilon_{mn}$ is singular. Indeed, let us suppose that the constraints are all first class, in which case there will exist $N$ independent null vectors $U^m_{(s)}$, i.e., $\epsilon_{mn} U^n_{(s)} = 0$. Then contracting $U^m_{(s)}$ with (\ref{epdelta}) it follows that the generator must satisfy the condition 
\beq
\frac{\partial A}{\partial y^m} U^m_{(s)} =0. \label{gammau}
\eeq
BG55 then showed that the commutator of two transformations satisfying this condition, with generators $A$ and $B$, has a generator
\beq
 \frac{\partial A}{\partial y^m} \delta_B y^m - \frac{\partial B}{\partial y^m} \delta_A y^m + \epsilon_{mn} \delta_A y^m \delta_B y^n =: \left\{A,B \right\}_{BG} \label{BG1}
\eeq
This is the definition of the Bergmann-Goldberg bracket. Most importantly, they proved that this commutator generator also satisfied the condition (\ref{gammau}), i.e, 
$$
\frac{\partial \left\{A,B \right\}_{BG}}{\partial y^m} U^m_{(s)} =0.
$$

The authors then wrote this expression in terms of the quasi-inverse $\eta^{mn}$ of the matrix $\epsilon_{mn}$ -  an object that we encountered earlier in a different context.. This has the property 
$$
\eta^{lm} \epsilon_{mn} = \delta^l_n - U^l_{(s)} V_n^{(s)},
$$
where $V_n^{(s)}$ is a null vector of $\eta^{ln}$. Thus multiplying (\ref{epdelta}) by the quasi-inverse there results
$$
\delta_A y^m = U^m_{(s)} V_n^{(s)} \delta y^n  -\eta^{mn} A_{,n}.
$$
Inserting this into (\ref{BG1}) gives, taking (\ref{gammau}) into account,
\beq
\left\{A,B    \right\}_{BG}  =  \eta^{mn} A_{,m} B_{,n}.
\eeq

We need one more consequence of our assumption that the constraints are first class, i.e., 
$$
\left\{C^a, C^b \right\} = c_{ab}^d C_d.
$$
Note that
$$
\left\{C^a,Y^\alpha \right\} \epsilon_{\alpha n} = \epsilon^{\mu \nu}C^a_{, \mu} Y^\alpha_{,\nu} \epsilon_{\rho \sigma} \zeta^\rho_{, \alpha} \zeta^\sigma_{, n}
= \epsilon^{\mu \nu} C^a_{, \mu}  \epsilon_{\rho \sigma}  \zeta^\sigma_{, n} \delta^\rho_\nu = \delta^\mu_\sigma C^a_{, \mu} \zeta^\sigma_{, n} = \delta^a_n =0,
$$
or, expanding the left hand side, we find that on the constraint hypersurface
$$
\left\{C^a, y^s \right\} \epsilon_{s n} =- \left\{C^a, C^b \right\} \epsilon_{b n} = -\left\{C^a, C^b \right\} c_{ab}^d C_d \epsilon_{b n} = 0.
$$
In other words, $\delta^a y^s := \left\{C^a, y^s \right\}$ is a null vector of $\epsilon_{s n}$. But, as we have seen, not only are the generators $A$ and $B$ invariant under this null transformation, but so is the commutator. The condition (\ref{gammau}) now has a clear physical meaning. It assumes the form
$$
0 = \frac{\partial A}{\partial y^m} \delta^a y^m = \left\{ A, C^a \right\}.
$$
In other words the permissible generators must be invariant under the action of the gauge group.
Thus we have the remarkable result that the Bergmann-Goldberg bracket gives the algebra satisfied by variables that are invariant under the action of the constraints. In the context of general relativity these invariants are constants of the motion - in addition to their independence of spatial coordinates! The algebra of observables of the reduced phase space is however not unique  because of the arbitrariness present in the quasi-inverse. A specific choice will result from the imposition of coordinate conditions. Indeed, when all the constraints are rendered second class in this manner, BG55 have proven that their generalized bracket becomes the Dirac bracket. There is another significant aspect of this result that has not received the attention it is due, perhaps in part because Bergmann and his collaborators have never stated  it explicitly. Or they might not have fully appreciated the significance of their result as is suggested by the Bergmann - Janis correspondence that will be discussed below. The fact is that one can obtain diffeomorphism invariants through the imposition of appropriate coordinate conditions - appropriate in the sense that they render the original set of first class constraints plus coordinate conditions second class. The proof follows from the authors' requirement that their generalized brackets are valid only for variables that are invariant under the action of the original first class generators. Before this paper appeared the argument had been made that simply by restricting the constrained phase space through coordinate conditions one was proceeding stepwise to the construction of invariants - although a direct demonstration that the variables satisfying the eventual Dirac bracket algebra were indeed invariant under the action of the diffeomorphism generators was lacking.

But on the other hand, to carry out the construction of the reduced algebra one must be in possession of invariants. Knowledge of the null vectors can serve in this search, as BG55 briefly illustrated with classical electromagnetism.
In this case the first class constraints are $\pi^0 = \frac{\partial {\cal L}}{\partial \dot A_0} = 0$ and 
$\frac{\partial \pi^a}{\partial x^a} = 0$ where $\pi^a = \frac{\delta {\cal L}}{\delta \dot A_a}$. This second constraint is the statement that the  longitudinal canonical momentum $\pi^a_l$ (the longitudinal electric field) is determined by the charged sources. The natural choice for the variables $y^m$ in this case is $y^m(\vec x) = \left(A_0, A_{l\,a}, A_{t\, b}, \pi_t^c\right)$, where the $t$ subscript represents the transverse field. As we have seen the constraints generate variations in null directions of $\epsilon_{mn}$, namely producing arbitrary variations in $A_0$ and $A_{l\, a}$. The resulting Bergmann-Goldberg algebra is therefore the algebra satisfied by the gauge invariant transverse electric and magnetic fields.

\section{Gravitational observables, coordinate conditions, and intrinsic coordinates}

Bergmann and his collaborators had therefore reached a definitive positive conclusion on the nature of the algebra of observables in classical general relativity, both in the Hamiltonian and Lagrangian approaches. Observables were invariant under the action of the full diffeomorphism group, and the Bergmann-Goldberg bracket represented their commutator algebra.  But there remained a problem. One still needed to find a complete set of invariant functionals that represented observables. Newman initiated an iterative construction of invariants in his 1956 thesis \cite{Newman:1956ab}, continued in a joint publication with Bergmann in 1957 \cite{Newman:1957aa}. But the evidence suggested that it might perhaps be helpful to try a new tack, modeled in part on success in constructing gauge invariants in electromagnetism as Bergmann explained in 1956 \cite{Bergmann:1956ab}. Bergmann's student Allen Janis completed a thesis in 1957 \cite{Janis:1957ab} in which he investigated how to find invariants through the imposition of coordinate conditions. The thesis confined attention to what he called `Lorentz type' conditions of the form 
\beq
C^j(y_A,y_{A,\rho}) = C^{*j}(y_A) + C^{jA \rho}(y_B)y_{A,\rho} = 0, \label{cj}
\eeq
 corresponding to the Lorenz gauge in electromagnetism.\footnote{A January 1957 APS abstract with Bergmann \cite{Janis:1957aa} is similarly limited in scope.} A 1958 joint publication \cite{Bergmann:1958aa} confined the analysis to coordinate conditions that do not depend on time derivatives $\dot y_A$. The idea in both instances is to add to the Lagrangian, as did Fermi in electrodynamics, a term that with suitable initial conditions renders the Lagrangian non-singular while imposing the gauge conditions. The appropriate term to add is $\frac{1}{2} a_{ij}C^i C^j$, and the initial data must be chosen consistent with (\ref{cj}). As in the previous papers, the focus here was to identify the temporal boundary terms that corresponded to the implementation of a canonical change in the physical state. The new consideration here was that infinitesimal changes needed to respect the coordinate conditions, i.e., neither the form of the conditions, nor their zero value (modulo the equations of motion) were permitted to change under variations $\bar \delta y_A$. They required that the equations of motion retained their form, and also that the Lagrangian might be altered by a total time derivative. This latter condition took an altered form since it was required to hold only for data that satisfied the coordinate condition. Therefore the requirement was that $\bar \delta L' = \dot Q + \frac{1}{2}  \delta a_{ij} C^i C^j$, where they reverted to the finite dimensional case. The generator boundary terms $C'^j$ again took the form $C' = - Q + \frac{\partial L'}{\partial \dot q_k}\bar \delta q_k$. And as in the earlier papers it was possible to show that $\dot C' + M^k \bar \delta q_k  = 0$ where $M^k = 0$ are the Euler-Lagrange equations. The $C'$ are therefore generally non-vanishing constants of the motion. They recognized that there might still remain some gauge freedom after the imposition of coordinate conditions, and these were to be factored out, just as the invariant transformations were factored out in forming the reduced algebra represented by the Bergmann-Goldberg algebra. Furthermore, and this was the message of the paper, these reduced algebras were identical, thereby showing that the imposition of coordinate conditions was legitimate. The equivalence was demonstrated explicitly for electromagnetism (in fact, focusing on only one Fourier mode of oscillation). They did express some misgivings about the practical use of this method as one was still  faced with the technical challenge of constructing constants of the motion whose corresponding transformations  respected the coordinate conditions. In closing they mention an alternative approach due to Komar and G\ea h\ea niau and Debever .
 
Komar had already in his 1956 thesis \cite{Komar:1956aa} (written under Wheeler's direction) used the G\ea h\ea niau and Debever \cite{Geheniau:1956aa} results as a means of distinguishing physically inequivalent solutions of Einstein's equations. Komar argued that in what he called `asymmetric' spacetimes, i.e., spacetimes that possess no Killing symmetries, the four independent Weyl scalars, formed with up to second derivatives of the metric, could be employed to define  what he called `intrinsic coordinates'. And in the frame in which these coordinates were employed, the metric components would be uniquely determined, and hence diffeomorphism invariants.\footnote{It is not clear when Bergmann became aware of Komar's work. G\ea h\ea niau summarized his joint work with Debever at the July 1955 Bern meeting.  They proved that there existed at most fourteen independent second order spacetime scalars.  Bergmann posed a question, following the presentation, relating to the existence of only four second differential order scalars that existed for vacuum spacetimes that possessed no symmetry. There is an edited proof (which does not appear to be in Bergmann's handwriting) of the  G\ea h\ea niau article that was  to appear in the Bern proceedings. The proof is in the Syracuse Bergmann archives (SUBA) in a folder labeled Bern Correspondence and does not mention Komar. The document states that Bergmann later had a private discussion with Wigner, and Wigner's response to Bergmann's inquiry follows. However, the ultimate published version contains the comment ``The question that must be decided ( and that Komar in Princeton has also addressed) concerned the characterization of definitively distinct solutions of Einstein's gravitational equations." ``Die Frage, die entschieden werden sollte (und die auch Herr Komar in Princeton angegriffen hatte) betraf die Charakterisierung wesentlich verschiedener Lösungen der EINSTEINschen Gravitationsgleichungen." It is likely that Bergmann heard the basis of his question directly from Komar at the April 1955 Washington meeting of the American Physical Society, where they were both present on the same day. Komar likely referred in his report \cite{Komar:1955ac} to an observation that would appear in his Ph. D. thesis \cite{Komar:1956aa}, citing the as yet unpublished \cite{Geheniau:1956aa} result that in a generic vacuum spacetime there exists four independent scalar invariants of  second differential order.  We know that Komar did go to Syracuse as a postdoctoral researcher, presumably at the beginning of the 1957 academic year. He stayed at Syracuse until 1963, promoted eventually to Associate Professor \cite{Goldberg:2013aa}.}  Komar even made a specific proposal for the coordinates, though without proof that his choices corresponded to spatial and timelike directions. Komar published a detailed proposal for the implementation of intrinsic coordinates in 1958 \cite{Komar:1958aa}, focusing first on the implications for the initial value problem in general relativity. He was able to show that the intrinsic coordinate choice led to unique temporal evolution. But of special concern to him was whether it would be possible to isolate from the redundant set of metric components and time derivatives a non-redundant set of four constants of the motion that would label physically distinguishable spacetimes. He addressed this question in studying the initial value problem.

Allen Janis, in an unpublished draft written in 1958\footnote{SUBA, Correspondence folder, {\it True observables and the generalized equivalence problem}}, did address the issue of the relationship in general between invariants and their associated algebra, and the variables that satisfied the Dirac algebra after the imposition of constraints. The former he called 'primary observables', and the latter 'secondary observables'. This proposed distinction elicited  a revealing discussion in correspondence in 1958 between Allen Janis and Bergmann. Bergmann's first remark concerns gauge freedom that might remain if a residual asymptotic invariance group is present, in which case he observes that an invariance group that `contains no arbitrary functions of \underline{four} coordinates" would be admissible. He also notes that in any case, as is true with the Komar approach, there will remain a redundance of observables. He then points to the recent work of Arnowitt, Deser, and Misner - to be published and also the work of Arnowitt and Deser reported ``at our meetings at Zurich and Neuchatel, who pointed out that the secondary observables can often be interpreted as primary observables, so that the distinction can often be more historical than actual. Take, as an example, the 'transverse' portion of some electromagnetic variable, e. g., the vector potential. One may either define the transverse vector potential as a certain functional of the vector potential (by means of an integral projection operator), so that it qualifies as a primary observable, without reference to a restricted coordinate frame; or one may introduce a special radiation gauge, in which case the transverse vector potential is simply the vector potential. Certainly, Komar's observables are interpretable either as primary or as secondary observables, depending on one's point of view."\footnote{SUBA, Correspondence folder, Letter from Bergmann to Janis dated September 15, 1958}

Correspondence by Bergmann with Dirac in 1959 sheds more light on the state of affairs with coordinate conditions at this time.\footnote{SUBA, Correspondence folder, Letter from Bergmann to Dirac dated October 9, 1959}  The letter was in response to Dirac's 1959 publication \cite{Dirac:1959aa} of a suggestion of suitable coordinate conditions in general relativity. Dirac had published in 1958 (D58) \cite{Dirac:1958ab} his groundbreaking paper in which he simplified the primary constraints in general relativity through the addition of a total divergence to the Lagrangian, thereby eliminating time derivatives of $g_{0 \mu}$. These components were thus freely prescribable. Bergmann wrote that ``(1) \ldots regardless of the motive in introducing the metric $g_{rs}$ on the initial hypersurface, the canonical transformation that you first published a year ago to simplify and kill the primary constraints, is both legitimate and successful. At this stage the total number of canonical field variables is reduced from twenty to twelve." He then goes on to discuss the proposed coordinate conditions. ``(2) What I have found most remarkable is the manner in which you have introduced coordinate conditions to change the secondary first-class constraints into second-class constraints, and eventually to reduce the theory to four canonical field variables. As far as I can see, your procedure and the one that Komar and I are still working on supplement each other: Your coordinates and field variables are intuitive, and results obtained will lend themselves to physical interpretation in terms of concepts with which we are familiar from other field theories. Your expansions will break down, or at least become unwieldy, if the field should be strong. You yourself have called attention to this fact. Komar and I work principally with closed-form expressions. Our variables are developed locally and require no solution of partial differential equations; but even if and when we complete our construction of the complete Lie algebra of our observables, including the dynamical laws, our results will be difficult to relate to more conventional concepts and procedures. Perhaps someone here will attempt to prove in some detail the mathematical equivalence of our two approaches. I do hope that we shall soon have another chance to compare notes on our progress personally, on either side of the Atlantic Ocean."\footnote{SUBA, Correspondence folder, Letter from Bergmann to Dirac dated October 9, 1959} 

An addendum to the same letter, dated October 14, raises another issue related to the problem of time. ``(3) When I discussed your paper at the Stevens conference yesterday, two more questions arose, which I should like to submit to you; To me it appeared that because you use the Hamiltonian constraint $H_L$ to eliminate one of the non-substantive field variables, $\kappa$, in the final formulation of the theory your Hamiltonian vanishes strongly, and hence all the final field variables, i.e., $\tilde e^{rs}$, $\tilde p^{rs}$, are `frozen' (constants of the motion). I should  not consider that as a source of embarrassment, but Jim Anderson says that in talking to you he found that you now look at the situation a bit differently. Can you enlighten me? If you have no objection, I should communicate your reply to Anderson and a few other participants in the discussion." 

Dirac responded on November 11, 1959\footnote{SUBA, Correspondence folder} first with the terse acknowledgement ``I fully agree with your comments (1) and (2)." Then he addressed the frozen time issue. ``If the conditions you introduce to fix the surface are such that only one surface satisfies the conditions, then the surface cannot move at all, the Hamiltonian will vanish strongly and all dynamical variables will be frozen. However, one may introduce conditions which allow an infinity of roughly parallel surfaces. The surface can then move with one degree of freedom and there must be one non-vanishing Hamiltonian that generates this motion. I believe my condition $g_{rs} p^{rs} \simeq 0$ is of this second type, or maybe it allows also a more general motion of the surface corresponding roughly to Lorentz transformations. The non-vanishing Hamiltonian one would get by substituting a divergence term from the density of the Hamiltonian." We are of course aware that Bergmann had much earlier concluded that observables must be independent of the coordinate time - and this would imply that the Hamiltonian resulting after the imposition of coordinate conditions would vanish identically, simply because the vanishing Hamiltonian constraint would be explicitly solved to express dependent degrees of freedom in terms of an independent set. This is a point that Anderson made in an undated letter to Bergmann commenting on a pre-publication draft he had obtained of Dirac's paper.\footnote{SUBA, Correspondence folder.} It turns out here that Dirac's coordinate condition did not fully eliminate the freedom in fixing the coordinate time. There remained a physically spurious one-parameter freedom. This issue was later partially addressed in Anderson's 1964 comparison of coordinate fixation techniques \cite{Anderson:1964ad}.\footnote{SUBA, Correspondence folder, This letter is also undated. Anderson writes ``Enclosed is what I hope is a corrected and correct version of the paper we discussed over the phone. As I mentioned then, I agree in the main with your comments and in fact appreciated them." Unfortunately we are not in possession of the earlier draft.}  It turns out that to fully eliminate this freedom the coordinate time must appear explicitly in a coordinate condition - as it does with Bergmann-Komar intrinsic coordinates.\footnote{See \cite{Pons:1997aa} for a proof.}

Indeed, while Bergmann and Komar were convinced that observables could not depend on coordinate time, they were also well aware that intrinsic time dependence was not excluded! They made this case quite explicitly in 1962 in their contribution to the Infeld Festschrift \cite{Bergmann:1962ad} in a section entitles `Time-dependent solutions'. They proposed interpreting an appropriately chosen function of canonical variables as the time. The idea was to choose a constant of the motion $C$, and find the canonical conjugate to  it, $\theta$. Then since the Poisson bracket of $C$ with the Hamiltonian $H$ weakly vanishes, $H$ must be independent of $\theta$ - at least on the constraint hypersurface. Thus one can solve for $C$, and write $C + h(\phi, \pi) = H = 0$, where the arguments of $h$ are the remaining phase variables with $C$ and $\theta$ excluded. Let $\{ , \}$ represent the original Poisson bracket, and $[ , ]$ the bracket formed with $\phi, \pi$. A variable $A$ will then satisfy the equation of motion
$$
\frac{d A}{d \theta} = \left\{ A ,H \right\} = \left[A ,h \right] + \left(\frac{\partial A}{\partial \theta}  \frac{\partial H}{\partial C}  - \frac{\partial C}{\partial \theta}  \frac{\partial \theta}{\partial C}\right) =  \left[A ,h \right]  + \frac{\partial A}{\partial \theta}.
$$

 \section{Dirac variables and Bergmann's interpretation}
 
 I have already mentioned, in conjunction with AB51, the puzzle regarding the failure to implement the full diffeomorphism Lie algebra. In fact, although AB51 did not remark on this fact, the mystery deepened with the deduction in this article that the associated Poisson bracket  algebra did not arise for the wide class of generally covariant field theories they considered. This broad class included the quadratic Einstein model  (\ref{EL}) although they did not display the explicit form for the secondary constraints for this specific case. They did however construct the general secondary constraints, and they displayed in their equation (7.7) the full algebra of these among themselves and also with the primary constraints - without calling attention to the Lie algebra puzzle. Bergmann's resolution was inspired by Dirac's gravitational Hamiltonian paper D58. One can best appreciate its significance in citing a passage from Bergmann's letter to Nathan Rosen in 1973, in which he proposes that Dirac be invited to talk at the 7th General Relativity and Gravitation conference to be held in Tel Aviv. He writes ``Having through an extended period wrestled with the same problems that he succeeded in solving - a viable Hamiltonian version of general relativity, I have the profoundest respect for his genius, second only (in my personal experience) to Einstein."\footnote{SUBA, Correspondence folder, Letter from Bergmann to Rosen, September 26, 1973, BA} I had not fully appreciated until recently, after rereading his {\it Handbuch der Physik} article, that Bergmann is not referring here to Dirac's earlier constrained Hamiltonian dynamics algorithm, developed concurrently with Bergmann, but to Dirac's apparently unwitting solution of the algebra conundrum. 
 
As noted above, Dirac was the first to publish a simplified version of the primary constraints in general relativity in which they appear as vanishing momenta \cite{Dirac:1958aa}. DeWitt had already reported a similar result at a Stevens meeting, although for the parameterized theory. This had inspired Anderson to seek an analogous result in the unparameterized model which he published shortly after Dirac \cite{Anderson:1958ac}. With the addition of an appropriate divergence to the Einstein Lagrangian Dirac showed that the four momenta conjugate to $g_{0\mu}$ must vanish. It follows immediately that the $g_{0\mu}$ are arbitrary. He was also the first to make the critical discovery that the canonical variables $g_{ab}$ and $\pi^{cd}$ were invariant under diffeomorphisms that left the spatial coordinates on a given spacelike hypersurface of constant $x^0$ fixed. In his 1962 {\it Handbuch de Physik} article \cite{Bergmann:1962ac}, Bergmann called these variables `D-invariant' in Dirac's honor. Under an arbitrary infinitesimal diffeomorphism $x'^\mu = x^\mu+ \xi^\mu(x)$, the change in D-invariants by definition does not depend on time derivatives of the $\xi^\mu$. Although Dirac did not demonstrate this explicitly, he clearly knew that given any vector $A_\mu$, in addition to $A_a$ constituting a D-invariant, there is a fourth D-invariant $A_L:= A_\mu n^\mu$, where $n^\mu =   - g^{0 \mu}\left(-g^{00}\right)^{-1/2} = \left(N^{-1}, -N^{-1} N^a \right)$ is the unit normal to the $x^0 = const$ hypersurface.\footnote{ Bergmann later in 1989, \cite{Bergmann:1989aa}, p. 298, characterized Dirac's procedure as having ``first appeared by magic".  It is possible that he might have been inspired to employ the normal by Weiss's work \cite{Weiss:1936aa} - which he nominally supervised. But I now doubt this. The Weiss construction employs the covariant normal. It does not depend on the metric.   But most significantly his canonical momenta are simply the conjugates of the field derivatives with respect to the parameter time, and his formalism did not contemplate reparameterization covariance - the context in which a notion of D-invariance would arise.} Dirac astutely reasoned that the time rate of change of the phase space variables would correspond to the naught component of a vector, namely, the variation of a variable $\eta$ under a change in time would be of the form
\beq
\delta \eta = \int d^3x \left\{ \mf{H}_0, \eta \right\} \delta t, \label{deltaeta}
\eeq
and the task was then to express $ \mf{H}_0$ in terms of D-invariants using $\mf{H}_L := \mf{H}_\mu n^\mu$. We read off that $\mf{H}_0 = \frac{1}{n^0} \left(\mf{H}_L - \mf{H}_a n^a \right) = N \mf{H}_L  + N^a  \mf{H}_a$.

Bergmann realized that this expression for the Hamiltonian could be obtained from an object $\int d^3 x \left(\mf{H}_L \epsilon^0 + \mf{H}_a \epsilon^a \right)$ that generates variations of D-invariants corresponding to infinitesimal coordinate transformations -  provided that these transformations involved a metric field dependence of the form 
\beq
\xi^\mu = \delta^\mu_a \epsilon^a+ n^\mu \epsilon^0, \label{modxi}
\eeq
since if one sets $\xi^\mu = \delta^\mu_0 \delta t$ in this expression it follows that $\epsilon^0 = N$ and $\epsilon^a = N^a$, and therefore the change in any variable $\eta$ under time evolution is given by (\ref{deltaeta}).
 As far as I am aware Bergmann was the first to note and publish this observation.\footnote{\cite{Bergmann:1962ac}, equation (27.11)} The implications are of course profound, for as Bergmann showed, this dependence eliminated the higher time derivatives that appear in the standard Lie algebra, as discussed previously. Indeed,  Bergmann derived the modified commutator of two infinitesimal transformations of the form (\ref{modxi}), obtaining
 \beq
 \epsilon^\rho = \delta^\rho_a \left(\epsilon^a_{1,b} \epsilon^b_2 - \epsilon^a_{2,b} \epsilon^b_1 \right) + e^{ab} \left(\epsilon^0_1 \epsilon^0_{2,b} - \epsilon^0_2 \epsilon^0_{1,b}\right) + n^\rho \left(\epsilon^a_2 \epsilon^0_{1,a} - \epsilon^a_1 \epsilon^0_{2,a}\right). \label{newep}
 \eeq
 Note that no time derivatives of the descriptors appear in this expression. Thus the problem described in section (\ref{two}) with the original diffeomorphism Lie algebra is resolved. 
Strangely, Bergmann did not give the corresponding Poisson brackets of the generators $\mf{H}_L$ and $\mf{H}_a$ although it is straightforward to derive them\footnote{He and Komar give the derivation later in \cite{Bergmann:1972aa}} and he must surely have known this bracket algebra which is now known as the Dirac algebra - even though Dirac apparently never published it. Higgs actually gives a partial result \cite{Higgs:1958aa,Higgs:1959aa}, but the offending bracket 
\beq
\left\{\mf{H}_L , \mf{H}'_L \right\} = \int d^3 x'' \mf{H}_a(x'') e^{ab}(x'') \left(\delta^3(x-x'') + \delta^3(x'-x'')\right) \frac{\partial}{\partial x^b} \delta^3 (x-x') \label{hlhl}
\eeq
 is missing. The first appearance in print I have found is in DeWitt \cite{DeWitt:1967aa}, equation (4.26a). The $e^{ab}$ represents the inverse of the 3-metric $g_{ab}$, and its appearance signifies that the group now constitutes exclusively a transformation group in phase space - even though every element can be associated with a specific four-dimensional general coordinate transformation, with none excluded.\footnote{In \cite{Pons:1997aa} we call it the `diffeomorphism related group'.}

\section{Conclusions} 

I have focused in this essay on the implications of the general covariance symmetry of Einstein's general theory of relativity as they were progressively investigated by Peter G. Bergmann and his collaborators at Syracuse University from 1949 to 1962. Bergmann believed that this underlying symmetry needed to be taken into account in the passage from the classical theory to an eventual quantum theory of gravity - in essence because it implied that spacetime coordinates could not of themselves carry physical information. In his mind the situation was similar to that in quantum electrodynamics where physical observables needed to be invariant under the action of the $U(1)$ gauge symmetry group. The situation in general relativity was of course enormously complicated by the fact that the gauge symmetry transformations of the metric field were themselves born of a corresponding transformation of spacetime coordinates - yet these very coordinates in the classical context not only served as identifying spatial labels but also tracked the field evolution in time. In fact, the underlying unity of classical spacetime would render this distinction between space and time labels as contrary to the principle of general relativity. Spatial field labels would under a four dimensional diffeomorphism transform into combined spatial and temporal indicators. 

We can discern a clear evolution in the Syracuse group's grasp and tentative resolution of the technical challenge they faced in attempting to implement this symmetry. It seemed natural that if the starting point was to be Einstein's theory one would need a procedure for reinterpreting the metric field and its spacetime derivatives as quantum operators. They first exploited identities that arose amongst classical field and velocity variables as a consequence of general covariance. The initial intent was to then perform a Legendre transformation to phase space and pursue a canonical quantization approach, to convert Poisson bracket relations to quantum commutators. The task was complicated by the appearance of primary constraining relations among these variables since they would lead to quantum inconsistencies, and there did not exist a procedure analogous to those employed in quantum electrodynamics that could be employed to eliminate them. It seemed preferable to work exclusively with variable functionals that were invariant under the action in phase space of the diffeomorphism group. An early attempt was made to formulate a quantum Schwinger-like action principle that would yield quantum commutators among field configuration and velocity operators, and its failure strengthened the case for first constructing classical invariants and then quantizing. The central objective was then to develop a phase space formalism that would incorporate general canonical transformations that might or might not correspond to a change in spacetime coordinates and in the process identify the gauge symmetry subgroup. Both  invariant and general infinitesimal transformations were revealed in the classical temporal boundary terms that have served for finite dimensional dimensional (non-field theoretic) systems as the foundation principle for Hamilton-Jacobi approaches. As had Rosenfeld much earlier (but for tetrad gravity), the Bergmann group showed that these latter vanishing terms terms would serve as the canonical generators of four-dimensional diffeomorphisms. The projection from configuration-velocity to phase space necessitated a procedure for finding the general solution of linear equations involving the singular Legendre matrix. The Bergmann school method differed mainly in outward appearance from that employed earlier by Rosenfeld. Dirac's roughly contemporaneous method was conceptually distinct and operationally somewhat simpler. And as we have seen Pirani and Schild succeeded in exploiting it to publish the first gravitational Hamiltonian - albeit using Bergmann's parameterized model. And Bergmann's student Penfield obtained the first non-parameterized Hamiltonian only shortly later. It should be stressed, however, that neither Pirani Schild nor Dirac himself ever concerned themselves with the realization of diffeomorphism symmetry as a canonical transformation group. 

We should also be cognizant of the fact that everything I have summarized so far occurred before the putative start of the Renaissance of general relativity in 1955. Major advances were made already in 1930, and one could contend that there did exist a degree of continuous progress in quantum gravity in the intervening period - at least in the period commencing in 1948.\footnote{See \cite{Blum:2018ad} for further documentation. See also \cite{Rickles:2019aa} for a magisterial analysis of this early period.}

The next chapter in the Syracuse story concerns the construction of classical gravitational invariants. These invariants would satisfy a Poisson bracket algebra that would transform between physically distinct solutions of Einstein's equations. The numerical values  of the invariants would indeed fix equivalence classes under the action of the diffeomorphism group, and the algebra could be understood as representing the factor group of the group of canonical transformations modulo the group of diffeomorphisms. The Bergmann-Goldberg construction represented this algebra, but it essentially required knowledge of invariants. Newman made some progress in an iterative construction. Another alternative, explored by Janis, was to impose appropriate coordinate conditions. But perhaps the most attractive possibility  - to me, at least -  was proposed by Komar and then jointly investigated with Bergmann. It was to use the classical geometry itself to locate spacetime landmarks that could be used as intrinsic spacetime coordinates. All metric components relative to this intrinsic coordinate system would be invariants. Yet this possibility was not fully exploited, and I speculate that this might be related to Dirac's delineation of new gravitational variables in 1958. On the one hand he argued that what we now know as the gravitational lapse and shift variables ought to be simply eliminated as canonical phase space variables since their evolution in time was arbitrary. And on the other hand, and perhaps even more importantly, in 1962 \cite{Bergmann:1962ac} Bergmann interpreted Dirac's gravitational Hamiltonian as reflecting the fact that his diffeomorphism generator actually produced canonical variations that corresponded to compulsory metric-dependent diffeomorphisms - the decomposition of infinitesimal diffeomorphisms into true 3-dimensional diffeomorphisms, and metric-dependent diffeomorphisms in the direction perpendicular to the temporal foliation. Both of these advances brought into question whether the generator constructed by Anderson and Bergmann in 1951 actually preserved the full four-dimensional diffeomorphism symmetry. This may explain why in the late 1960's Bergmann and Komar began to look closely at a Hamilton-Jacobi approach to observables in general relativity - a chapter in the Syracuse saga that will appear soon \cite{Salisbury:2020ac}.  But they did continue to seek a group theoretical interpretation of Dirac's canonical generator, with a groundbreaking paper BK72 \cite{Bergmann:1972aa}. Referring to D58, Bergmann observed in 1979 that  ``At the time of the Dirac papers the nature of the commutators that he constructed was not entirely clear. Had Dirac merely discovered a new Lie algebra, or was his Lie algebra the germ of a group? If so, what was the nature of the group? In 1971 A. Komar and I were able to answer that question. There was a new group."\footnote{\cite{Bergmann:1972aa}, p. 175}
 They showed that the generator was that of a transformation group of the metric variables -  including the 3-metric. This came about because of the appearance of the  inverse three-metric $e^{ab}$ in (\ref{newep}). In fact, as a consequence of the spatial derivatives - reflected also in the descriptor algebra (\ref{newep}) - derivatives of this metric arose at increasingly higher order with nested commutators. BK72 concluded that the metric dependence was spatially non-local. Bergmann stressed the compulsory metric dependence in a later essay \cite{Bergmann:1979aa} in which he referred to the `fading world point'. Each individual symmetry mapping, viewed as a functional of the 3-metric, ``sends a given point into a cloud of points, depending on the particular Cauchy data involved \ldots What this whole analysis may teach us is that the world point by itself possesses no physical reality."\footnote{\cite{Bergmann:1972aa}, p. 176}

We witness in this period in the early 1960's the emergence of a competing quantum gravitational formalism - Wheeler's geometrodynamics.  He was inspired by the Feynman path integral approach in flat spacetime in which the action served as a quantum phase. In his vision a quantum transition amplitude between initial and final temporal states played a primary role. Although he did recognize that the theory must take into account the three-dimensional spatial diffeomorphism symmetry, he believed that the full four dimensional symmetry was lost. It is replaced by what he called `multi-fingered time', a notion that first appears in print in his 1967 Battelle lectures \cite{Wheeler:1968aa}. There was a related substantial debate, beginning in 1960,  involving Wheeler, his student Sharp, Misner, Bergmann, and Komar over the so-called thin sandwich conjecture in which Wheeler claimed that the 3-geometry already carried information about time.\footnote{See \cite{Salisbury:2020ac} for a detailed analysis.} Arnowitt, Deser, and Misner  \cite{Arnowitt:1962aa} (ADM) of course invented their own first-order gravitational formalism in this same period, and with a similar disavowal of four-dimensional covariance. Their view paralleled that of Dirac in this regard, and in fact their ultimate theory is equivalent to Dirac's, as interpreted by Bergmann, whereby the contemplated infinitesimal diffeomorphims are understood as undergoing a perpendicular decomposition.\footnote{Some authors do dispute this equivalence. See in particular \cite{Kiriushcheva:2011aa}.} The upshot for all was that the fourth so-called scalar constraint needed of course to be imposed, but there was no clear relation to the original four-dimensional diffeomorphism symmetry. Rather, it acquired a `dynamical interpretation'.\footnote{Bryce S. DeWitt Papers, 1919, 1946-2007, and undated, Archives of American Mathematics, Dolph Briscoe Center for American History, University of Texas at Austin, Research status report for period 1 December 1964 to 31 May, 1965, Box 4RM235}  The reference is to Bryce DeWitt's suggestion in the early 1960's to Wheeler that the constraint could be implemented in a Hamilton-Jacobi formalism, in a form that is now known as the Wheeler-DeWiit equation. This formalism has since the 1960's overtaken the field, and the reasons for its dominance certainly merit a careful historical analysis. Granted, related to its presumptive correctness is the companion so called problem of time which continues to occupy the minds of physicists and philosophers. I can perhaps not render an objective judgement as I and my collaborators carry on the Bergmann tradition, but in my opinion the term renaissance does not apply to quantum general relativity. Rather, what we encounter, in my opinion, is more analogous to the pre-Renaissance loss of classical scholarship. 

\section*{Acknowledgements}
I would like to thank J\"urgen Renn and the Max Planck Institute for the History of Science for support offered me as a Visiting Scholar. Thanks also to Alexander Blum for his critical reading and many constructive suggestions.

\bibliographystyle{plain}
\bibliography{qgrav-V19}

\end{document}